\documentclass[]{rQUF2e}
\usepackage{graphicx}
\usepackage{color}
\usepackage{amsfonts}
\usepackage{srcltx}
\bibpunct{(}{)}{,}{a}{}{,}

\begin{document}


\title{Impact of meta-order in the Minority Game}

\author{
A C Barato$\dag$, I Mastromatteo${\ddag}$, M Bardoscia$^{\ast}$${\dag}$\thanks{$^\ast$Corresponding author. Email: marco.bardoscia@ictp.it } and M Marsili${\dag}$
\\\vspace{12pt}
\normalfont{$\dag$ The Abdus Salam International Centre for Theoretical Physics, Trieste 34014, Italy\\
$\ddag$ International School for Advanced Studies, Trieste 34136, Italy}
\\\vspace{12pt}
}

\maketitle

\begin{abstract}
We study the market impact of a meta-order in the framework of the Minority Game. This amounts to studying the response of the market when introducing a trader who buys or sells a fixed amount $h$ for a finite time $T$. This perturbation introduces statistical arbitrages that traders exploit by adapting their trading strategies. The market impact depends on the nature of the stationary state: we find that the permanent impact is zero in the unpredictable (information efficient) phase, while in the predictable phase it is non-zero and grows linearly with the size of the meta-order. This establishes a quantitative link between information efficiency and trading efficiency (i.e. market impact). 
By using statistical mechanics methods for disordered systems, we are able to fully characterize the response in the predictable phase, to relate execution cost to response functions and obtain exact results for the permanent impact.
\end{abstract}

\begin{keywords}
Minority Game; Market Impact; Meta-order; Market Microstructure.
\end{keywords}

\section{Introduction}
Market impact is the effect on the price caused by an order, and it measures the tendency of prices to move up (down) subsequent to a buy (sell) trade of a given size \citep{Bouchaud:2010dk}. Understanding market impact is a fundamental problem in financial markets, and it has recently been the subject of several studies, both empirical \citep{Hasbrouck:2001zr, Plerou:2002ve, Lyons:2002ly, Lillo:2003vn, Almgren:2005xh, Moro:2009wy, Toth:2011sj} and theoretical \citep{Torre:1997qf, Grinold:2000bh, Farmer:2005fk, Gabaix:2005dq, Bouchaud:2008pb, Toth:2011bs}. From a theoretical perspective, market impact is related to the problem of understanding how information is incorporated into security prices \citep{Bouchaud:2008pb, Kyle:1985zh, Glosten:1985kx, Hasbrouck:1991iw}.

For practitioners, controlling market impact is essential in order to limit execution costs, which arise because each trade impacts adversely on the price. In addition, traders often face the problem of executing orders whose volume is much larger than the outstanding liquidity which is available for trading at each moment in time, which is typically a factor $10^{-3}$ smaller than the volumes exchanged daily \citep{Bouchaud:2008pb}. A common strategy consists in splitting large orders  in a sequence of child orders that are executed incrementally. This sequence of trades is commonly called \emph{meta-order}. The trading activity generated by meta-orders has been argued to be one of the origins of long range auto-correlation in the flow of orders in financial markets \citep{Bouchaud:2008pb}, and generates market impact which extends beyond that of single orders.
One robust empirical finding is the fact that market impact is a concave function of the order size both for individual orders \citep{Bouchaud:2008pb} and for meta-orders \citep{Moro:2009wy, Toth:2011bs}, in contrast with early models \citep[e.g.][]{Kyle:1985zh} which predict a linear impact. 

Understanding market impact requires addressing the strategic interaction between liquidity providers and liquidity takers, and in particular modeling how the former exploit the predictable trading patterns of the latter. Kyle's seminal paper \citep{Kyle:1985zh} and its most recent extension to meta-orders \citep{Farmer:2011} build on this key ingredient, but are limited to simple (representative) agent settings. In addition, the assumption of perfectly informed agents hardly suits to realistic settings\footnote{Typically, liquidity takers are perfectly informed on how much they need to trade, but not on the future price of assets.}. More recent theoretical studies on market impact have been done within the framework of order-book models \citep{slanina08}, which provide an accurate description of the market microstructure. We take a complementary approach by using a coarse-grained description of the market, focusing on the collective behavior of a set of heterogeneous adaptive market participants.

The natural framework for addressing this issue is the Minority Game (MG) \citep{challet97, Challet:2005wl, coolen05}, which is a paradigmatic model of heterogeneous adaptive agents who repeatedly interact in a market setting using simple ``trading strategies'' to either buy or sell an asset, with the objective of being in the minority side (i.e.\ selling when the majority buys or vice-versa). In spite of its simplicity, the MG offers a picture of a financial market as a many-agent interacting system where speculators ``predate'' on predictable patterns introduced by other agents, much as in real financial markets. As such, the MG shows, in a simplified setting, how markets aggregate information as a result of speculative activity. In brief, one can distinguish an ``asymmetric'' phase, characterized by the presence of statistical arbitrages when speculative pressure is low, from a ``symmetric'' phase, when there are many speculators and the market is unpredicable \citep{challet99}. Close to the transition between the two regimes several stylized facts observed in financial markets can be successfully described by slightly generalizing the original MG problem \citep{Challet:2001cr, bouchaud01, Challet:2003nx}.

The MG is the appropriate setting for studying the market impact of meta-orders, because it reproduces the response of 
adaptive traders to perturbations introduced by the persistent activity of meta-orders.  A meta-order is introduced in the MG by adding  an extra agent that consistently buys (or sells) for a finite time period, after which the extra agent is removed, and the meta-order finishes. In order to analyze how this perturbation affects the price, one needs to make an assumption on how the price is related to excess demand. Here we assume a market clearing mechanism \citep{Marsili:2001mc} taking a linear relation between them. Alternative approaches are discussed in Sec.\ \ref{sectionConclusion}. This allows us to estimate the impact of the meta-order on price, during and after its execution. Our main findings are: \emph{i)} in the predictable phase the permanent impact is nonzero, while in the unpredictable phase there is no permanent impact; \emph{ii)}  the permanent impact can be computed analytically, relating it to response functions. Finally, \emph{iii)} we find that the market impact is linear with the size of the order, as a consequence of the assumed linear relation between price and excess demand.

The organization of the paper is as follows. In the next section we define the model, discuss its phase transition and explain how the meta-order is introduced. In Sec.\ \ref{sectionResults} we present our results and in Sec.\ \ref{sectionConclusion} we draw the main conclusions. Furthermore, we add appendices with the calculations used to obtain the analytical results.

\section{Grand-canonical Minority Game}
\label{sectionGCMG}

\subsection{Model definition}
The MG describes a set of agents interacting in a market over a sequence of periods labeled by a discrete time variable $t \in \mathbb{N}$. Each period corresponds to the trading activity in a time window $[t,t+\Delta t)$. A realistic value for $\Delta t$ must be much larger than inter-trade times, e.g.\ $\Delta t\gg 10$ s, and can be thought as being the typical time over which meaningful information accumulates, e.g. few minutes\footnote{One method to estimate time-scales over which information flows is that of studying the dynamics of cross-correlations between different stock prices. On very liquid stocks, \citet{Mastromatteo:2011ab} show that for $\Delta t\le 1$ min price dynamics is dominated by noise, whereas \citet{Borghesi:2005cd} show that the structure of correlations of daily returns can be recovered already from the analysis of returns over $\Delta t = 5$ minutes. Similar results have been discussed in \citet{toth-kertesz}.}.
Speculators act in the market by using trading strategies that process the value of a public information signal $\mu$ and submit orders accordingly. In practice, a number of trading strategies is constantly being evaluated and the most profitable are used. 
Profitable strategies are those that dictate to buy when the majority is selling or vice-versa, i.e.\ that place the trader in the minority side as frequently as possible. This captures in a simple manner the market making activity of liquidity providers.

Formally, we consider the grand-canonical MG setup \citep{Challet:2005wl}. The information to which traders react can take $P$ values, labeled by $\mu=1,\ldots,P$. Every time step a new signal $\mu(t)$ is drawn at random. We suppose there are two types of traders: liquidity takers -- also called producers in \citet{Challet:2005wl} -- and speculators or liquidity providers. There are $N_p$ producers and $N_s$ speculators and their only difference is that while the former react to information signals in a deterministic manner, the latter are adaptive. 

Specifically, each agent $i$ is endowed with a trading strategy $\{a_i^{\mu}\}_{\mu=1}^P$, that dictates whether to buy ($a_{i}^{\mu}=+1$) or to sell ($a_i^{\mu}=-1$) when public information is $\mu(t)=\mu$. These strategies are independently drawn for each player, uniformly in the set of all $2^{P}$ possible binary functions $f:~\{1,\ldots,P\}\to \{\pm 1\}$. However, while producers always trade according to their strategy, speculators may choose to trade or not, adapting their behavior to the profitability of their strategy. This choice is encoded in a dynamical variable $\phi_i(t)$, which takes the value $\phi_i(t)=1$ if agent $i$ decides to trade, and $\phi_i(t)=0$ otherwise. The profitability of a strategy is quantified by a score $U_i(t)$ that is updated according to
\begin{equation}
U_i(t+1)=U_i(t)-a_{i}^{\mu(t)}A(t) \, ,
\label{defscore}
\end{equation}  
where the excess demand $A(t)$ is given by
\begin{equation}
A(t)= A_p^{\mu(t)}+\sum_{i=1}^{N_s}a_{i}^{\mu(t)}\phi_i(t)
\label{defA}
\end{equation}
and $A_p^\mu$ is the sum of the contributions from producers when $\mu(t)=\mu$. The decision of playing or not is taken in accordance to the score of the speculator. In the simplest case, agent $i$ does not trade if $U_i(t)<0$ whereas she trades if the score is positive\footnote{This rule can be generalized using a stochastic choice model \citep{Challet:2005wl}, without changing the main results presented in this paper.}. In summary, a time-step iteration of the model is realized in the following way:
\begin{itemize}
\item[-] randomly generate a information variable $\mu(t) \in \{1,\ldots,P\}$;
\item[-] for each speculator $i$, set $\phi_i(t)=1$ if $U_i(t)\geq 0$, and $\phi_i(t)=0$ otherwise;
\item[-] calculate $A(t)$ using Eq.\ (\ref{defA}); 
\item[-] for each speculator $i$, update $U_i(t+1)$ by Eq.\ (\ref{defscore});
\item[-] set $t\to t+1$.
\end{itemize}

\subsection{Phase transition in MG}
As discussed in \citet{Challet:2003nx} this model has a very rich phenomenology. In the limit of large markets, analog to the thermodynamic limit in physics, the non-trivial regime arises when $P,N_s,N_p\to \infty$, while $n_s=N_s/P$ and $n_p=N_p/P$ remain finite. A central role in the characterization of the collective behavior is played by the quantity
\begin{equation}
H= \frac1P\sum_{\mu=1}^{P}\langle A|\mu\rangle^2 \, , 
\label{pred}
\end{equation}
where 
\begin{equation}
\langle A|\mu\rangle=A_p^\mu+\sum_{i=1}^{N_s}a_{i}^{\mu}\langle\phi_i\rangle
\end{equation}
denotes the average of $A(t)$ conditional on $\mu(t)=\mu$. $H$ detects the presence of statistical arbitrages, in that if $H>0$ there is at least one value of $\mu$ which allows for a statistical prediction of the excess demand when $\mu(t)=\mu$. For this reason $H$ is called predictability. A remarkable feature of the MG is that the dynamics converges to states of minimal predictability  $H$ \citep{Challet:2005wl}. This is not only appealing conceptually, as it says that speculators leave the market as unpredictable as possible given their strategies, but it also opens the possibility for analytic approaches to the MG. Indeed, it turns the study of the stationary state of the MG into the analysis of the minima of $H$, as a function of the variables $\langle \phi_i\rangle$. This is akin to studying the ground state properties of a disordered system in physics, a problem for which a plethora of powerful tools has been developed in recent years \citep{Challet:2005wl, coolen05}. In Appendix \ref{ap1} we show how this is done by calculating the critical line and the predictability in the asymmetric phase. In brief, one finds that
\begin{equation}
\label{eqHdef}
\frac{H}{N_s}=\frac{n_p+Gn_s}{(1+\chi)^2} \, ,
\end{equation}
where $G$ is given in Eq.\ (\ref{eqG}) in the Appendix \ref{ap1} and 
\begin{equation}
\label{eqdefchi}
\chi=\frac{1}{N_s}\sum_{i=1}^{N_s} \frac{\delta \langle\phi_i\rangle}{\delta a_i}
\end{equation}
can be interpreted as an agent susceptibility, as it measures the response of agents' behavior to a change $a_i^\mu\to a_i^\mu+\delta a_i$ in their strategies\footnote{This is a standard result \citep{parisi-mezard-virasoro}, see e.g. Appendix A1.1 in \citet{fwmd} for an explicit derivation in an interacting agents' context. In the dynamic theory of MG \citep{coolen05}, $\chi$ emerges as the integrated response of an instantaneous perturbation. Loosely speaking, it measures the time over which agents's behavior maintains memory of perturbations.}.
The relation between $\chi$ and $G$ on $n_s$ and $n_p$ is discussed in the appendix, here we focus on the resulting picture of the behavior of the MG, which can be summarized as follows. For a fixed value of $n_p$, if $n_s$ is small enough the market is predictable ($H>0$) and the susceptibility $\chi$ is finite. In words, the number of speculators is not enough to exploit the amount of statistical arbitrages introduced by producers. When $n_s$ increases, $H$ decreases whereas $\chi$ increases, in agreement with Eq.\ (\ref{eqHdef}). Beyond a critical value $n_s^*$ the susceptibility diverges ($\chi=\infty$) and the market is unpredictable ($H=0$). The point $n_s^*$ marks a phase transition in that it separates two regions with distinctly different features. In this paper we focus on the case $n_p=1$, for which the critical value is $n_s^*\approx4.15$.

\subsection{Meta-orders in the Minority Game}
In order to introduce a meta-order in the grand-canonical MG we add a fixed buyer to the system. That is, we add one extra producer that always buys independently of the public information pattern. This extra producer buys a quantity $h$ of assets at each time, for $T$ time steps. This models a meta-order of duration $T$, and total size $hT$. Since typical relaxation times in the MG are proportional to $P$, we'll consider $T$ proportional to $P$ in what follows, which might correspond to time-scales of one trading day. Note that turning on the meta-order corresponds to changing the excess demand to
\begin{equation}
A_h(t) = h\theta(T-t)\theta(t)+ A_p^{\mu(t)}+\sum_{i=1}^{N_s}a_{i}^{\mu(t)}\phi_i(t) \, ,
\label{Aperturbed}
\end{equation}
where $\theta(x)=1$ for $x>0$ and $\theta(x)=0$ otherwise.
This way of modeling the unwinding of a meta-order is reminiscent of a very common scheduling strategy (the TWAP scheme), in which the trader's target is to execute a quantity of shares
which grows linearly in time \citep{berkowitz-logue-noser}.

It can be shown that, imposing a market clearing condition \citep[see][]{Marsili:2001mc, De-Martino:2006os}, the quantity $A(t)$ is proportional to the return of the logarithm of the price, which is given by
\begin{equation}
\log p(t+1) = \log p(t)+\frac 1P A(t) \, .
\end{equation}
Different market mechanisms can imply a different relation between price change $p(t+1) / p(t)$ and excess demand $A(t)$. This relation has implications in how the market impact depends on the order size, which is of central importance in market impact studies. Our definition is likely to be appropriate at large time scales (say, between one day and a week), in which one expects stochasticity
not to play a dominant role in the description of the market clearing mechanism.
We shall return to this point in the conclusions. We define the price impact as the averaged log-price change in presence of the meta-order, minus the trend component which would have been present in the unperturbed case. This corresponds to the quantity
\begin{equation}
\Delta(t)= \frac 1P\sum_{0 \leq s<t} \overline{\Big< A_h(s) -  A_{h=0}(s) \Big>} \, ,
\label{eqimpact}
\end{equation}
where $s=0$ corresponds to the beginning of the meta-order, the average over the stationary state distribution is denoted with $\langle \cdots \rangle$ and the one over the possible choices of the fixed strategies\footnote{The strategies are fixed in the sense that they do not change during the whole evolution of the system, i.e.\ before the beginning of the order, during the order and after its completion. In physical jargon such variables would be called \emph{quenched} strategies. However, for the sake of clarity, we will continue to refer to them as fixed strategies.} $a_{i}^{\mu}$ has been written as $\overline{(\cdots)}$. This averaging procedure mimics the averages in empirical studies, which are on different orders at different times, as well as on different markets \citep{Moro:2009wy, Toth:2011sj, Toth:2011bs}.
Notice that $\langle A_{h=0}\rangle$ is expected to be zero only after averaging over the fixed strategies, as in the asymmetric phase its value is finite, and fluctuates from realization to realization.\footnote{It can be estimated \citep{Challet:2005wl} that $\langle A_{h=0} \rangle$ is finite for any realization of the MG, and it is described by a Gaussian random variable of mean zero and variance $H/N_s$.}
The averages in Eq.\ (\ref{eqimpact}) are numerically estimated according to the following procedure. Initially all the scores are set to zero and the model is run until it relaxes to the stationary state. In this initial phase of the evolution speculators adapt to the information injected by producers in the absence of a meta-order. Then we compute $\langle A_{h=0} \rangle$ taking a time average in the stationary state\footnote{The average has to be taken on times much longer than those over which we observe the impact $\Delta(t)$.}. After that the meta-order is turned on and we start measuring the impact. Finally the average over the fixed strategies is computed by repeating this procedure for a large number of realizations. 

We will also consider $\Delta(t)$ for $t>T$, since we are interested in the response of the system after the meta-order is turned-off. Of particular interest is the permanent impact, which reads $\Delta^*=\lim_{t\to \infty}\Delta(t)$, and the average execution cost
\begin{equation}
p(0) \bar\Delta = p(0) \left[ \frac{1}{T}\sum_{t=1}^T \Delta(t) \right]
\end{equation}
which measures the mean price increment incurred during the meta-order. In particular, it is a simple exercise to show that the cost of an execution for a trader is related to the maximum
price increment $p(0) \Delta (T)$ as
\begin{equation}
\label{defaec}
\frac{\bar \Delta}{\Delta(T)} = \frac{\overline{ \left< \frac{1}{T} \sum_t p(t) - p(0) \right>} }{\overline{ \left< p(T)  - p(0)\right> }} \, .
\end{equation}
The estimation of the above quantity provides a practical mean to evaluate the concavity of price impact: for a market impact of the form $\Delta (t) \propto t^\alpha$, one finds that $\bar \Delta /\Delta(T) = 1/(1+\alpha)$, which indicates that concave impact functions have to be associated with a value of $\bar \Delta /\Delta(T) > 1/2$, while for convex impact functions a value smaller than $1/2$ has to be expected.

\section{Results}
\label{sectionResults}
\begin{figure}
\centering
\includegraphics[width=73mm]{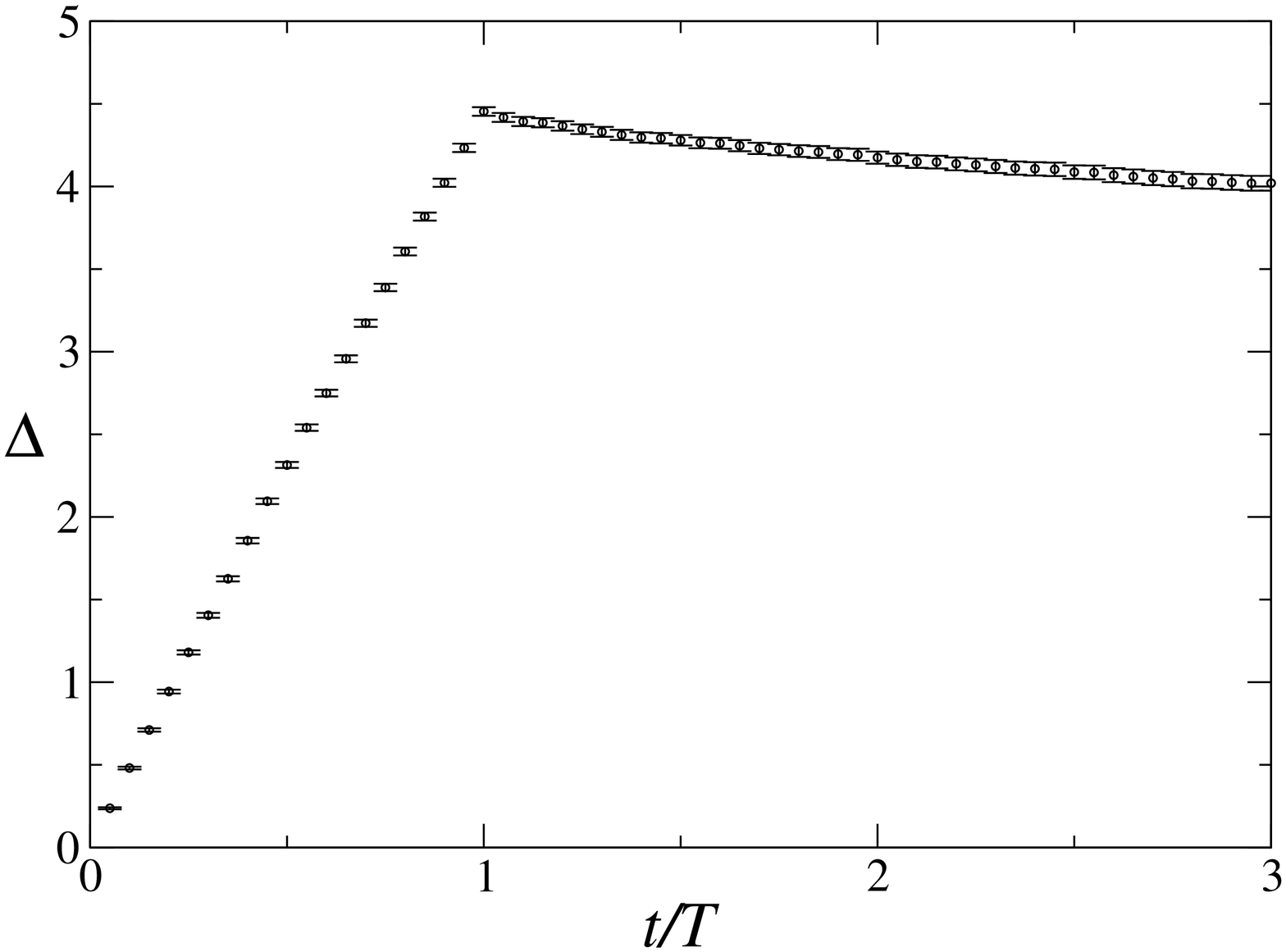}
\includegraphics[width=73mm]{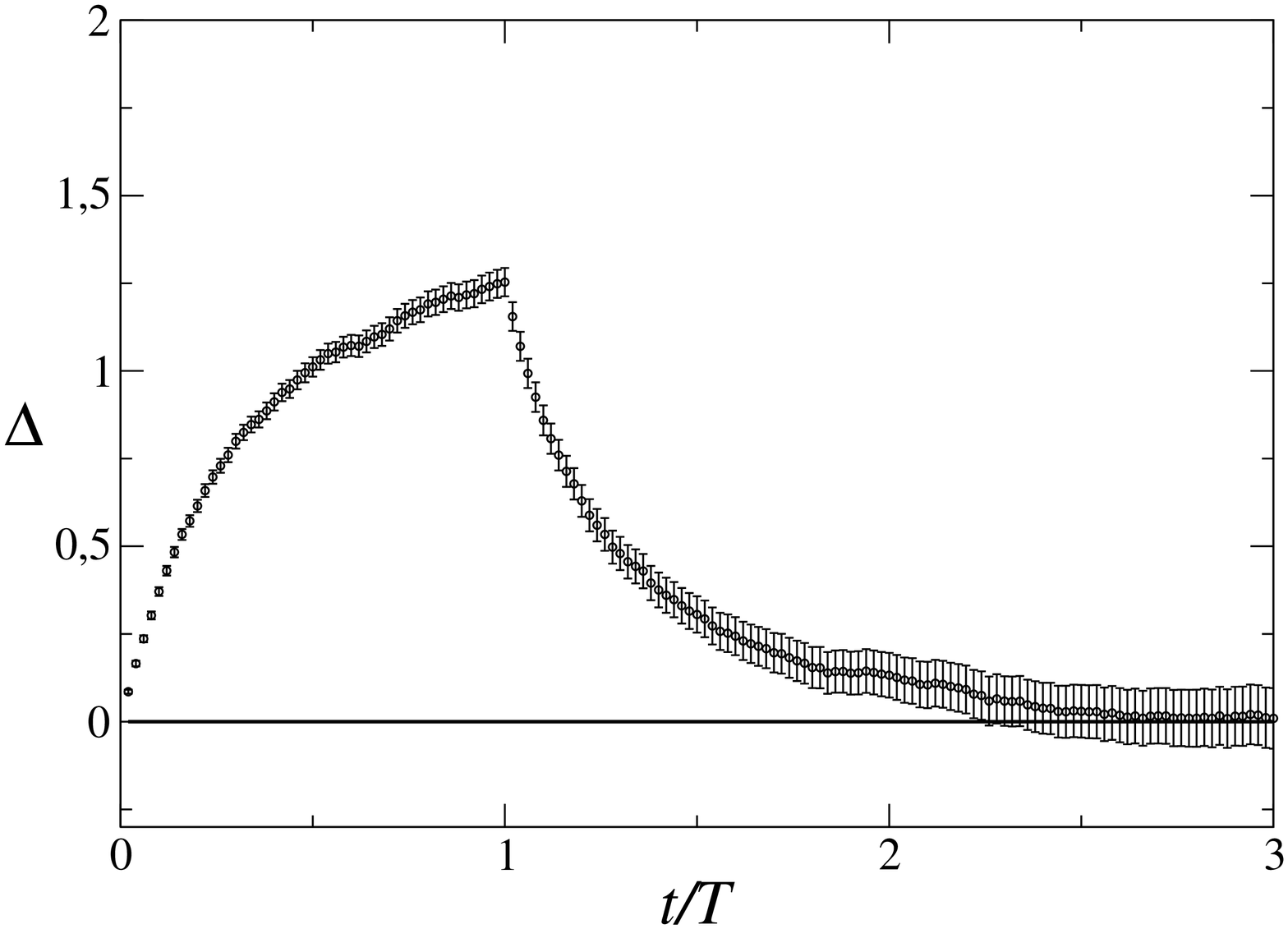}
\caption{Comparison between the MG in the asymmetric and symmetric phases for $P=400$, $T=5P$, and $5\times10^3$ realizations. On the left side we have $n_s=1$ where the predictability and the permanent impact are non-zero. On the right side $n_s=5$, where $H=0$. 
}
\label{fig1}	
\end{figure}

Let us first analyze how the impact changes depending  on whether the model is in the symmetric or the asymmetric phase. In Fig.\ \ref{fig1} we plot the impact $\Delta(t)$, obtained from numerical simulations, as a function of time for $n_s=1$ (predictable phase) and $n_s=5$ (unpredictable phase). In the predictable phase the impact grows linearly during the meta-order. After the meta-order finishes, it decays to a constant faster than a power law, but with a long relaxation time. In the unpredictable phase the situation is rather different. During the meta-order the impact grows initially very fast and then it reaches a saturation value. After the meta-order is finished the impact quickly relaxes back to zero. Therefore, as a main result of this paper, we find that if the market is unpredictable ($H=0$) the permanent impact is zero and if the market is predictable ($H>0$) the permanent impact is non-zero. We point out that the permanent impact is zero in the symmetric phase as long as the the duration of the meta-order $T$ is big enough such that the impact saturates before the end of the meta-order. If $T$ is much smaller  than the time for saturation to set in, the permanent impact in the symmetric phase is non-zero.   

In terms of the excess demand we have the following picture: immediately after the order is turned on, the (average) excess demand jumps from $0$ to $h$. During the execution of the meta-order it relaxes to a value smaller than $h$. In the unpredictable phase, this value is zero, whereas in the predictable phase it is finite, implying a linear growth of the impact. At the end of the meta-order ($t=T$) the excess demand suddenly drops by $h$, thus becoming negative. Finally it relaxes back to zero, regardless of the phase. 

This result can be explained in the following way. If predictability is zero, it means that the market is information free and the system is able to absorb the extra information of the extra fixed buyer. Therefore, the value of the excess demand relaxes back to zero during the meta-order. On the other hand, when the market is predictable the number of speculators is not enough to consume all the information available. Hence, when the perturbation is added, the system reaches a new stationary state where the value of the excess demand is non-zero.   

\begin{figure}
\centering
\includegraphics[width=73mm]{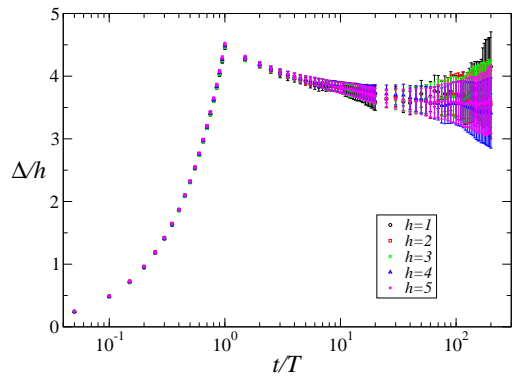}
\caption{Market impact as a function of time for different values of $h$, for  $n_s=1$, $P=400$, $T=5P$, and $5\times10^3$ realizations. The collapse of the curves shows that the market impact grows linearly with h.}
\label{fig2}	
\end{figure}

Fig.\ \ref{fig2} shows that the impact grows linearly with the order size $Q=hT$. Indeed the rescaled impact $\Delta/h $ collapses on the same curve, when plotted against $t/T$. This is a consequence of the assumed market mechanism and of linear response theory, as we shall see, and it contrasts with results of empirical \citep[e.g.][]{Toth:2011bs} which find a square root law $\Delta\sim\sqrt{Q}$. We will come back to this issue in the concluding section.

\begin{figure}
\centering
\includegraphics[width=73mm]{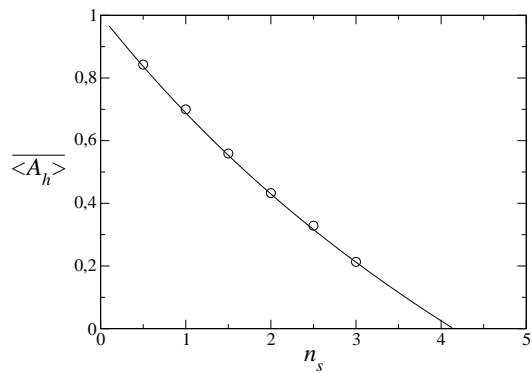}
\caption{Comparison between  the saturation value of the aggregate in the presence of the meta-order $\overline{\langle A_h\rangle}$ obtained analytically (full line) and from numerical simulations (open circles). The error on the points from numerical simulations is at most of the size of the open circles.}
\label{fig3}	
\end{figure}

These results are fully consistent with the theoretical predictions. 
As we show in Appendix \ref{ap2}, the stationary state of the system in the presence of the meta-order can be calculated by minimizing the predictability $H$ with the modified excess demand (\ref{Aperturbed}). This allows us to calculate the saturation value of the excess demand before the meta-order is finished, which is given by 
\begin{equation}
\label{avgAh}
\overline{\langle A_h \rangle} = \overline{\langle A_{h=0}\rangle}+\frac{h}{1+\chi} \, ,
\end{equation}
In order to obtain it numerically we calculate the slope of the curves of $\Delta(t)$ when the linear growth of the impact sets in, after some initial transient. In Fig.\ \ref{fig3} we can see that the numerical results are in full agreement with the exact calculation. 

It is worth noticing that $\overline{\langle A_h \rangle}/h$, which measures the response of the market to a perturbation $A\to A+h$, is inversely proportional to the susceptibility $\chi$ of agents' behavior. The market impact is larger when agents are not very susceptible to perturbations, whereas it is minimal in the symmetric phase ($H=0$), when agents respond strongly to any perturbation. 

\begin{figure}
\centering
\includegraphics[width=73mm]{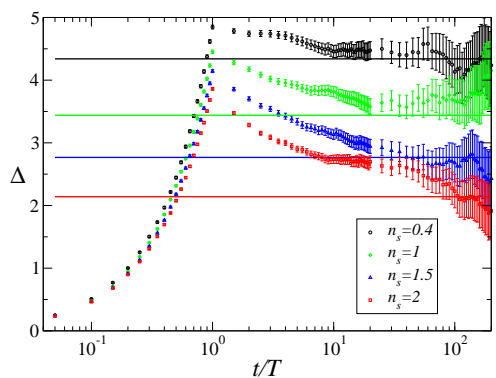}
\caption{Comparison of numerical simulations with the permanent impact given by formula (\ref{permanent}). The numerical results were obtained with $P=400$, $T=5P$, and $5\times 10^3$ independent realizations.
The straight horizontal lines indicate the value of the permanent impact obtained from (\ref{permanent}).
}
\label{fig4}	
\end{figure}

In order to address the dynamic effects of market impact, it is necessary to reformulate the dynamics in a continuum time description which acknowledges the fact that characteristic  times in the MG are of order $P$ \citep{Challet:2005wl,coolen05}. As long as the perturbation $h$ is much smaller than $A \sim \sqrt{P}$, linear response theory holds, i.e.,
\begin{equation}
\overline{\langle A_h(t) \rangle} = \overline{\langle A_{h=0}(t) \rangle}+ \frac{1}{1+\chi} \int_{-\infty}^{t/P} d\tau \; K(t/P-\tau) \tilde h(\tau) \, ,
\label{linear_resp}
\end{equation}
where $\tilde h(\tau) = h(P \tau)= h \theta(T/P-\tau)\theta(\tau)$ is the perturbation and $K(\tau)$ is a kernel that contains the relaxation dynamics. Note that if $t,T\to\infty$ we expect to recover the stationary value of Eq.\ (\ref{avgAh}), hence $\int_{-\infty}^{\infty} d\tau K(\tau) = 1$. Also causality requires $K(\tau)=0$ for $\tau<0$.
We remark that in order to reproduce the small time behavior of the model (i.e.\ $\langle A_h(t) - A_{h=0}(t) \rangle \approx h$ for small $t$), the function $K(\tau)$ must have the form
\begin{equation}
\label{kernel}
K(\tau) = (1+ \chi) \delta(\tau) - \chi K_r(\tau) \, ,
\end{equation}
where $K_r(\tau)$ is non-singular in zero, causal ($K_r(\tau)=0$ for $\tau<0$) and $\int_{-\infty}^{\infty} d\tau K_r(\tau) = 1$.
$K_r(t)$ in Eq.\ (\ref{kernel}) captures how adaptive traders detect the statistical arbitrage introduced by the meta-order and react to it, by removing it as a result of their activity.
If in addition we assume that $K_r(\tau)$ decays fast enough when $\tau\to\infty$, then within linear response theory, we show in Appendix \ref{ap3} that the permanent impact is 
\begin{equation}
\Delta^*(h,T)= \frac{hT}{P(1+\chi)} \, .
\label{permanent}
\end{equation}
The assumption on $K(s)$ is fully consistent with simulations, where the excess demand is found to relax faster than a power-law, which is expected since we are not considering the model at criticality. In Fig.\ \ref{fig4} we see agreement between Eq.\ (\ref{permanent}) and numerical results. However, relaxation times are rather long even for values of $n_s$ considerably far from the critical value $n_s^*\approx 4.15$, making the error bars become very large before $\Delta$ is saturated. 

Finally, the average execution cost, which is derived in Appendix \ref{ap4}, is given by
\begin{equation}
\label{aec}
\frac{\bar\Delta}{\Delta(T)} = 1 - \frac{1}{2}\left( \frac{\kappa_T^0 - \kappa_T^2}{\kappa_T^0 - \kappa_T^1} \right) \, ,\qquad \kappa^m_T = \int_0^1 dx \,x^m K(x T / P) \, .
\end{equation}
For a kernel $K(\tau)$ of the form (\ref{kernel}) and execution times $T/P$ much smaller than the characteristic relaxation time of $K_r(\tau)$, this quantity is well approximated by
\begin{equation}
\label{smallT}
\frac{\bar\Delta}{\Delta(T)} \approx \frac{1}{2} + \frac{T}{12\,P } \left( \frac{\chi}{1+\chi} \right) K_r (0) \, ,
\end{equation}
while at large times we obtain a value $\bar\Delta / \Delta(T) = 1/2$ which signals a linear impact.
This should be compared with $\bar\Delta/\Delta(T) \approx 0.6\div 0.7$ reported in empirical studies \citep[e.g.][]{Farmer:2011,Toth:2011bs}, and with the arguments in \citet{Toth:2011bs}.

\section{Conclusion}
\label{sectionConclusion}
The market impact has been the subject of intense discussion recently. 
Proposals of universal laws have been put forward \citep{Farmer:2011,Toth:2011bs} on the basis of empirical analysis and simple models. Here we address this issue within the framework of the Minority Game. At odds with zero-intelligence (or $\epsilon$-intelligence \citep{Toth:2011bs}) models, the MG retains a level of rationality of the agents without making strong assumptions on their information on future returns \citep[as in][]{Farmer:2011}. Rather, traders learn adaptively on the profitability of their strategies. In addition, the minority rule is rather appropriate if one focuses on market making activity, where the profit of agents comes from matching imbalances in demand. Finally, the MG is the ideal platform to address how markets digest information on order imbalances from meta-orders and to address the issue of the market impact beyond its immediate consequences. 

We have shown that the MG provides realistic curves for the market impacts and sheds light on the mechanisms which are responsible for it. Most interesting, we derive a relation between market impact and information efficiency, whereby the permanent impact of a meta-order is smaller the more the market is free of statistical arbitrages. For a perfectly ``symmetric'' market where returns are statistically unpredictable, prices should revert to the values before the order started, and hence the permanent impact should be zero. Understanding whether this relation extends to other modeling frameworks or whether it is confirmed by empirical data, is a very interesting issue for future research.

A noticeable shortcoming of our analysis pertains the prediction of a linear impact with order size ($Q=Th$ in our notations), which contradicts the observed concave behavior $\Delta\sim Q^\gamma$, with $\gamma\in[0.5,0.8]$. It is easy to identify the origin of this behavior within the MG, in the assumed market mechanism. This assumes a matching of orders at a single deterministic price, fixed by market clearing. Indeed, the price itself does not enter into the definition of the MG. Strictly speaking the MG describes the dynamics of the excess demand, and how this is determined by market makers who adaptively respond to it, in their effort to match it dynamically. Therefore, on one side this suggest to look for alternative definitions of the price in the MG, 
capturing more faithfully the dynamics of the order book. On the other, this suggest to look at empirical measures of the market impact in terms of the demand imbalance itself. 

The analysis of the more realistic case in which the size of the orders $hT$ is Pareto distributed would also be a relevant extension of the present work, and could allow the model to describe persistence of the order flow (see \citet{Farmer:2011} and \citet{Toth:2011bs}). There are two possible choices: either $h$ can be Pareto distributed and $T$ fixed, or, vice versa, $T$ is Pareto distributed and $h$ fixed. In the former case it would not be possible to apply linear response theory, since there will be realizations with $h \simeq P$, which is not a small perturbation. In the latter case it must be considered that when one measures the market impact averaged over orders of different volumes for a given value of volume $Q$, only the orders whose volume is larger than or equal to $Q$ are taken into account. If the difference in volume depends only on the difference in the length of the order then, at a given time, only the orders which are still standing are taken into the average. As a consequence, it is easy to show that, at least during the execution of the meta-order, the market impact is still linear with respect to the volume. Alternatively, it would be very interesting to study a Pareto distributed order flow for the environment (i.e. by assigning to the producers strategies which lead to large fluctuations of the excess demand $A_p^\mu$). In this case the market stationary properties should be different, and an anomalous response to a meta-order could emerge. It is not easy to anticipate what the behavior of the system would be as the problem becomes harder to solve analytically.

Finally, as discussed in \citet{Toth:2011bs}, the market impact is expected to cross-over to a linear behavior for times longer than few days. This supports the view that the MG provides a description of market activity on intermediate time-scales, with $P$ representing a time-scale between one day and a week.

\section*{Acknowledgements}
We gratefully acknowledge discussions with J.-P. Bouchaud, F. Lillo, R. Mantegna, B. Toth, and M.A. Virasoro.

\appendices

\section{The stationary state of the grand-canonical Minority Game with replica symmetric approach}
\label{ap1}
The problem of calculating the macroscopic observables of the grand-canonical MG in the stationary state can be solved by finding the minimum (also referred as ground state) of a Hamiltonian function, which is precisely the Lyapunov function minimized by the dynamics of the grand-canonical MG \citep{De-Martino:2006os}
\begin{equation}
H_0(\mathbf{m}) = \frac{1}{2N_s} \sum_{\mu=1}^P \langle A | \mu \rangle^2 \, ,
\end{equation}
where
\begin{equation}
\langle A | \mu \rangle = \sqrt{N_p} x_\mu+ \sum_{i=1}^{N_s} a_{i}^{\mu} m_i \, 
\end{equation}
and $\mathbf{m}=(m_1,m_2,\ldots,m_{N_s})$. Each of the $m_i\in[0,1]$ is the average value of $\phi_i$ in the stationary state, and $x_\mu$ is a Gaussian variable with zero mean and unit variance. For convenience we are considering an Hamiltonian that is different from the predictability $H=2n_sH_0$. Therefore, in order to recover the free energy associated with $H$, one has simply to multiply the free energy obtained with $H_0$ by $2n_s$.    

Let us define  the partition function
\begin{equation}
Z=  \textrm{Tr}_{\mathbf{m}}\exp(-\beta H_0)=\textrm{ Tr}_{\mathbf{m}}\left<\exp \left(\sum_{\mu=1}^Pi\sqrt{\frac{\beta}{N_s}} \langle A | \mu \rangle \, z_\mu\right)\right>_{z},
\end{equation}
where we have used the Hubbard-Stratonovich trick in the second equality and 
\begin{equation}
\left<f(z)\right>_{z}= \frac{1}{\sqrt{2\pi}}\int_{-\infty}^{\infty}\exp(-z^2/2)f(z)dz \, .
\end{equation}
The minimum of $H_0$ averaged over the fixed strategies $a_{i}^{\mu}$ can be obtained from the partition function in the limit of $\beta\to\infty$. It is given by
\begin{equation}
\lim_{N_s \to \infty}\min_{\scriptsize\textbf{m}}\frac{\overline{H_0}}{N_s} = -\lim_{N_s\to \infty}\lim_{\beta\to \infty}\lim_{n\to 0}\frac{1}{\beta N_s n}\ln \overline{Z^n} \, ,
\end{equation}
where the over-bar denotes an average over the fixed strategies and the variable $n$ was introduced in order to calculate the average $\overline{\ln Z}$ through the one of $\overline{Z^n}$, a strategy known as replica trick \citep{parisi-mezard-virasoro}. We now proceed calculating $\overline{Z^n}$.

It is a simple exercise \citep[see][]{De-Martino:2006os} to show that after averaging out the fixed strategies we obtain
\begin{equation}
\overline{Z^n}= \textrm{Tr}_{\{\mathbf{m}_a\}}\left<\exp\left[-\frac{\beta}{2}\sum_{a,b}z_a^\mu z_b^\mu\left(\frac{n_p}{n_s}+\frac{1}{N_s}\sum_{i}m_{ia}m_{ib}\right)\right]\right>_z \, ,
\label{disorderout}
\end{equation} 
where the indices $a,b$ run over the $n$ replicas.

Introducing the overlap matrix $G_{ab}= \frac{1}{N_s} \sum_i m_{ia} m_{ib}$ and the response matrix $\tilde{G}_{ab}$ \cite[see][for details]{De-Martino:2006os}, it follows that
\begin{equation}
\overline{Z^n}= \int d\mathbf{G} \, d\mathbf{\tilde{G}}\exp(-\beta n N_s f(\mathbf{G},\mathbf{\tilde{G}})) \, , 
\end{equation}
where
\begin{eqnarray}
f(\mathbf{G},\mathbf{\tilde{G}}) & = & \frac{1}{2 \beta n_s n}\textrm{Tr}\ln\left[\mathbf{I}+\beta\left(\frac{n_p}{n_s}+\mathbf{G}\right)\right]
+ \frac{\beta}{2 n_s n} \sum_{a \leq b} G_{ab} \tilde{G}_{ab}\nonumber\\
& \phantom{=} & - \frac{1}{\beta n}\ln \textrm{Tr}_{\{m_a\}}\exp\left(\frac{\beta^2}{2 n_s} \sum_{a\leq b}\tilde{G}_{ab}  m_a m_b \right) \, .
\end{eqnarray}

We now take the ansatz (known as the replica symmetric ansatz) $G_{ab} = g + (G-g) \delta_{ab}$ and $\tilde{G}_{ab}= 2R-(R+\rho/\beta)\delta_{ab}$. Note that in terms of the original variables
\begin{equation}
\label{eqG}
G=\frac{1}{N_s}\sum_{i=1}^{N_s} \langle\phi_i\rangle^2 \, .
\end{equation}
In this way the free energy $f(\mathbf{G},\mathbf{\tilde{G}})$ becomes
\begin{eqnarray}
f(G,g,R,\rho) & = & \frac{1}{2\beta n_s}\ln\left(1+\beta(G-g)\right)+\frac{1}{2n_s}\frac{n_p/n_s+g}{1+\beta(G-g)}+\frac{R}{2n_s}\beta(G-g)\nonumber\\
& \phantom{=} & -\frac{1}{\beta}\left<\ln\int_0^1dm\exp(-\beta V(m,z))\right>_z \, ,
\label{free1}
\end{eqnarray}
where $V(m,z)= \frac{1}{2n_s}\rho m^2+\sqrt{\frac{R}{n_s}}zm$. In the limit of $N\to\infty$ the free energy is dominated by its minimum and the saddle-point equations are
\begin{eqnarray}
G&=&F(\zeta)\nonumber\\
\frac{\chi}{1+\chi}&=&\frac{n_s}{2} \, \textrm{Erf}(\zeta/\sqrt{2})\nonumber\\
\rho&=&\frac{1}{1+\chi}\nonumber \\
R&=&\frac{n_p/n_s+G}{(1+\chi)^2} \, ,
\end{eqnarray}
where $\chi=\beta(G-g)$, $\zeta=\rho/\sqrt{Rn_s}$, and $F(\zeta)= -\frac{1}{\sqrt{2\pi}\zeta}\exp(-\zeta^2/2)+\frac{1}{2\zeta^2} \textrm{Erf}(\zeta/\sqrt{2})+\frac{1}{2} \textrm{Erfc}(\zeta/\sqrt{2})$. Rearranging these equations we obtain the following recipe to calculate the macroscopic observables. First one should solve the following transcendental equation in order to obtain $\zeta$ as a function of the control parameters $n_s$ and $n_p$,
\begin{equation}
n_sF(\zeta)= \frac{1}{\zeta^2}-n_p \, .
\label{eqnum}
\end{equation}
After that, one can express the macroscopic observables as function of $\zeta$ by using the saddle point equations. For example, it is easy to calculate the minimum of the free energy to obtain the predictability as
\begin{equation}
H= 2  n_s^2f_{SP}\,P=N_s \left( n_p+n_s F(\zeta) \right) \left( 1 - \frac{n_s}{2}\textrm{Erf}(\zeta/\sqrt{2}) \right)^2 \, ,
\end{equation} 
where $f_{SP}$ is the free energy at the saddle point. Moreover the susceptibility $\chi$ is written as
\begin{equation}
\chi= \frac{n_s\textrm{Erf}(\zeta/\sqrt{2})}{2 - n_s \textrm{Erf}(\zeta/\sqrt{2})} \, .
\label{eqchi}
\end{equation}

\section{Perturbing the Hamiltonian}
\label{ap2}
We now consider the perturbed Hamiltonian
\begin{equation}
H_h(\mathbf{m}) = \frac{1}{2N_s} \sum_{\mu=1}^P \big(\langle A | \mu \rangle+h\big)^2 \, .
\end{equation}
This corresponds to the Lyapunov function that is minimized by the dynamics when the meta-order is turned-on. Therefore, it gives the stationary state measure for the case with an extra buyer.
 
The calculations in the present case are analogous to the calculations in Appendix \ref{ap1}. For example, with the perturbation, Eq.\ (\ref{disorderout}) changes to
\begin{equation}
 \overline{Z^n_h} =\left<\exp \left(  i \sqrt{\frac{\beta}{N_s}}\sum_{a,\mu}h z_\mu^a\right)
\textrm{ Tr}_{\{\mathbf{m}_a\}}\exp\left[-\frac{\beta}{2}\sum_{a,b}z_a^\mu z_b^\mu\left(\frac{n_p}{n_s}+\frac{1}{N_s}\sum_{i}m_{ia}m_{ib}\right)\right]
\right>_z \, ,
\end{equation}
where the $Z_h$ indicates the partition function associated with $H_h$. This leads to the following free energy
\begin{eqnarray}
f_h(\mathbf{G},\mathbf{\tilde{G}})=\frac{h^2 P }{2 N_s^2 n} \sum_{a,b} M^{-1}_{ab}+f(\mathbf{G},\mathbf{\tilde{G}}) \, ,
\end{eqnarray}
where $M_{ab}=\delta_{ab}+\beta(n_p/n_s+G_{ab})$ and $f(\mathbf{G},\mathbf{\tilde{G}})$ is the free energy given by (\ref{free1}). With the replica symmetric ansatz we have $\sum_{ab}M_{ab}^{-1}=\frac{n}{1+\beta(G-g)}+O(n^2)$, leading to
\begin{eqnarray}
f_h(G,g,R,\rho) = \frac{h^2 P}{2 N_s^2 (1 + \chi)} + f(G,g,R,\rho) \, .
\label{free2}
\end{eqnarray}
Hence, we have obtained the free energy for the perturbed case. Moreover, if we write
\begin{equation}
Z_h = \textrm{Tr}_{\mathbf{m}}\exp\left(-\frac{\beta}{2N_s}\sum_{\mu}\big(\langle A | \mu \rangle+h\big)^2\right) \, ,
\end{equation}
we get
\begin{equation}
\lim_{N_s \to \infty}   \lim_{\beta \to \infty} \left( - \frac{N_s}{\beta P} \right) \left( \frac{d \ln Z_h}{d h} \right) =  \left< \frac{1}{P}\sum_\mu\big(\langle A | \mu \rangle+h\big) \right> = \langle A_h \rangle \, .
\end{equation}
This gives
\begin{equation}
\overline{\langle A_h \rangle}= \frac{N_s^2}{P}\frac{d}{dh}f_h \, .
\end{equation}
Finally, from Eq.\ (\ref{free2}) we obtain Eq.\ (\ref{avgAh}), where $\chi$ is given by Eq.\ (\ref{eqchi}), after numerically solving Eq.\ (\ref{eqnum}).

\section{Permanent impact}
\label{ap3}
In the following we show how the permanent impact relates to the stationary state quantities of the model calculated above. We will consider the continuous time description of the dynamics \citep{Challet:2005wl} as justified in Sec.\ \ref{sectionResults}, and define accordingly the quantities $\tilde T  = T/P$ and
\begin{equation}
\tilde \Delta (\tau) = \Delta (\tau P) \approx  \int_0^{\tau} d\tau^\prime  \; \overline{\Big< A_h(\tau^\prime P ) \Big>  -  \Big<A_{h=0}(\tau^\prime P) \Big>} \, .
\end{equation}
Within linear response theory, Eq.\ (\ref{linear_resp}) yield the response at time $t$ of the system due to a perturbation $\tilde h(\tau)$, in terms of the kernel $K(\tau)$. Note that $\int_{-\infty}^{\infty} d\tau K(\tau) = 1$ and that causality requires $K(\tau) = 0$ for $\tau <0$. 
From Eq.\ (\ref{eqimpact}) the permanent impact $\Delta^*=\lim_{\tau \to\infty} \tilde \Delta(\tau)$ is given by
\begin{equation}
\Delta^* = \frac{1}{1+\chi} \int_0^\infty d\tau  \int_{-\infty}^{\tau} d\tau^\prime \; K(\tau -\tau^\prime) \,\tilde h(\tau^\prime) \; .
\end{equation}
Assuming that $K(\tau)$ is an integrable function, we have
\begin{equation}
K(\tau) = \int_{-\infty}^{+\infty} \frac{d\omega}{2\pi} K_\omega e^{-i \omega \tau} \, .
\end{equation}
Therefore, the  expression for the permanent impact can be written as 
\begin{eqnarray}
\Delta^* &=& \frac{h}{(1+\chi)} \int_0^\infty d\tau  \int_{0}^{\tilde T} d\tau^\prime  \; \int_{-\infty}^{+\infty} \frac{d\omega}{2\pi} K_\omega e^{-i \omega (\tau - \tau^\prime)}\nonumber \\
&=& \frac{h}{(1+\chi)} \lim_{\epsilon \to 0^+} \int_{-\infty}^{+\infty} \frac{d\omega}{2\pi} \left( \frac{1}{i\omega + \epsilon} \right) \left( \frac{1 - e^{(i\omega + \epsilon) \tilde T}}{-i\omega - \epsilon} \right)K_\omega \, .
\end{eqnarray}
The value of this integral is determined by the value of the residues of the integrand on the plane $\Im \omega > 0$, provided that $\left| K_\omega / \omega \right| \rightarrow 0$ for $| \omega | \rightarrow \infty$. If $K_\omega$ has no poles on such domain, which corresponds to the assumption of causality $K(\tau) =0$ for $\tau < 0$, the only contribution to the integral comes from a second-order pole in $\omega = i \epsilon$. Hence, using $\int_{-\infty}^{+\infty} d\tau K(\tau) = 1$ we obtain our final formula
\begin{equation}
\Delta^*= i \frac{h}{1+\chi}  \lim_{\epsilon \to 0^+} \left( -i K_{i\epsilon} \tilde T \right) = \frac{hT}{P(1+\chi)} \, .
\end{equation}

\section{Average execution cost}
\label{ap4}
We calculate in the following the expression for the average execution cost as defined in Eq.\ (\ref{defaec}), and compare it with the maximum price increment $\Delta(T) p(0)$. To calculate those
quantities it is sufficient to assume linear response by plugging Eq.\ (\ref{linear_resp}) into the definitions of $\Delta (T)$ and $\bar \Delta$. 
The former quantity can be obtained straightforwardly integrating by parts, and results
\begin{eqnarray}
\Delta(T) & \approx & \frac{h}{1+\chi} \int_0^{\tilde T} d\tau \int_0^\tau d\tau^\prime  K(\tau - \tau^\prime) \nonumber \\
 & = &  \frac{h}{1+\chi} \left[ \int_0^{\tilde T} d\tau \, (\tilde T - \tau)  \, K(\tau) \right] \nonumber \\
 & = &  \frac{h\tilde T^2}{1+\chi} \Big[ \kappa_T^0 - \kappa_T^1 \Big] \, ,
 \end{eqnarray}
 where the averages $\kappa_T^m$ have been defined in Eq.\ (\ref{aec}).
The expression of the average execution cost is analogously derived by using integration by parts
\begin{eqnarray}
\bar\Delta & = & \frac{1}{T}\sum_{1\leq t < T} \Delta(t) \approx \frac{1}{\tilde T} \int_0^{\tilde T} \! d\tau \, \tilde \Delta(\tau) \nonumber \\
 & = &  \frac{h\tilde T^2}{2(1+\chi)} \Big[ \kappa_T^0 - 2 \kappa_T^1 + \kappa_T^2 \Big] \, .
 \end{eqnarray}
 By taking the ratio of those expressions one obtains Eq.\ (\ref{aec}).
 The functions $\kappa^m_T$ can formally be computed by inserting the expression (\ref{kernel}) into their definition. In particular we obtain the series expansion
 \begin{equation}
 \kappa_T^m = \delta_{m=0} \, (1+\chi) \, \tilde T^{-1}  - \chi \sum_{n=0}^\infty \frac{K^{(n)}_r(0)}{n!}\frac{\tilde T^n}{n+m+1} \, ,
 \end{equation}
where $K_r^{(n)}(0)$ is the $n$-th derivative of the function $K_r(\tau)$ evaluated in zero. Such expansion allows to calculate Eq.\ (\ref{smallT}) which is obtained in the limit of small $\tilde T$.\footnote{
We stress that the presence of a delta function in the kernel $K(\tau)$ is crucial to obtain a non-convex impact in the small time limit, because for a function of the type $K(\tau) = K_r(\tau)$ one would
obtain $\bar \Delta / \Delta (T) \approx 1/3$ as a consequence of the scaling $\Delta (t) \sim t^2$.}
The execution cost in the limit of $\tilde T$ much larger than the characteristic relaxation time $\tau_r$ can be estimated by noting the scaling
\begin{equation}
 \kappa_T^m \sim  \frac{1}{\tau_r} \left( \frac{\tau_r}{\tilde T} \right)^{m+1} \, ,
\end{equation}
so that just the $ \kappa_T^0$ terms contributes to Eq.\ (\ref{aec}). This implies that in this limit $\bar \Delta / \Delta (T) $ tends to the value of $1/2$.


\begin{thebibliography}{99}

\bibitem[\protect\citeauthoryear{Almgren {\itshape{et al.}}}{2005}]{Almgren:2005xh} 
Almgren R., Thum C., Hauptmann E. and Li H., 
Direct estimation of equity market impact, 
{\em Risk}, 
2005, 
{\bf 18}, 5762.

\bibitem[\protect\citeauthoryear{Berkowitz {\itshape{et al.}}}{1988}]{berkowitz-logue-noser}
Berkowitz S., Logue D. and Noser E., 
The total cost of transacting on the NYSE, 
{\em J. Financ.}, 
1988, 
{\bf 41}, 97--112.

\bibitem[\protect\citeauthoryear{Borghesi {\itshape{et al.}}}{2007}]{Borghesi:2005cd} Borghesi C., Marsili M., Miccich\'e S., 
Emergence of time-horizon invariant correlation structure in financial returns by subtraction of the market mode, 
{\em Phys. Rev. E}, 
2007, 
{\bf 76}, 026104.

\bibitem[\protect\citeauthoryear{Bouchaud {\itshape{et al.}}}{2001}]{bouchaud01} 
Bouchaud J.-P., Giardina I. and M\'ezard M., 
On a universal mechanism for long-range volatility correlations, 
{\em Quant. Fin.}, 
2001, 
{\bf 1}, 212--216.

\bibitem[\protect\citeauthoryear{Bouchaud {\itshape{et al.}}}{2008}]{Bouchaud:2008pb} 
Bouchaud J.-P., Farmer J. D. and Lillo F.,
How Markets Slowly Digest Changes in Supply and Demand, 
{\em Handbook of Financial Markets: Dynamics and Evolution}, 
pp.~57--156, 
2008 
(Elsevier: Academic Press).

\bibitem[\protect\citeauthoryear{Bouchaud}{2010}]{Bouchaud:2010dk} 
Bouchaud J.-P., 
Price Impact. 
{\em Encyclopedia of Quantitative Finance}, 
2010 
(New York: John Wiley \& Sons).

\bibitem[\protect\citeauthoryear{Challet and Zhang}{1997}]{challet97} 
Challet D. and Zhang Y.-C., 
Emergence of cooperation and organization in an evolutionary game, 
{\em Physica A}, 
1997, 
{\bf 246}, 407--418.

\bibitem[\protect\citeauthoryear{Challet and Marsili}{1999}]{challet99} 
Challet D. and Marsili M., 
Phase transition and symmetry breaking in the minority game, 
{\em Phys. Rev. E}, 
1999, 
{\bf 60}, R6271.

\bibitem[\protect\citeauthoryear{Challet {\itshape{et al.}}}{2001}]{Challet:2001cr} 
Challet D., Chessa A., Marsili M. and Zhang Y.-C., 
From Minority Games to real markets, 
{\em Quant. Fin.}, 
2001, 
{\bf 1}, 168--176.

\bibitem[\protect\citeauthoryear{Challet and Marsili}{2003}]{Challet:2003nx} 
Challet D. and Marsili M., 
Criticality and market efficiency in a simple realistic model of the stock market, 
{\em Phys. Rev. E}, 
2003, 
{\bf 68}, 036132.

\bibitem[\protect\citeauthoryear{Challet {\itshape{et al.}}}{2005}]{Challet:2005wl} 
Challet D., Marsili M. and Zhang Y.-C., 
{\em Minority Games}, 
2005 
(Oxford: Oxford University Press).

\bibitem[\protect\citeauthoryear{Coolen}{2005}]{coolen05} 
Coolen A. C. C., 
{\em The mathematical theory of Minority Games}, 
2005 
(Oxford: Oxford University Press).

\bibitem[\protect\citeauthoryear{De Martino and Marsili {\itshape{et al.}}}{2006}]{De-Martino:2006os} 
De~Martino A. and Marsili M., 
Statistical mechanics of socio-economic systems with heterogeneous agents, 
{\em J. Phys. A},  
2006, 
{\bf 39}, R465.

\bibitem[\protect\citeauthoryear{Farmer {\itshape{et al.}}}{2005}]{Farmer:2005fk} 
Farmer J. D., Patelli P. and Zovko I., 
The predictive power of zero intelligence in financial markets, 
{\em PNAS}, 
2005, 
{\bf 102}, 2254--2259.

\bibitem[\protect\citeauthoryear{Farmer {\itshape{et al.}}}{2011}]{Farmer:2011} 
Farmer J. D., Gerig A., Lillo F. and Waelbroeck H., 
How efficiency shapes market impact, 
\texttt{arXiv:1102.5457}, 
2011.

\bibitem[\protect\citeauthoryear{Gabaix {\itshape{et al.}}}{2005}]{Gabaix:2005dq} 
Gabaix X., Gopikrishnan P., Plerou V. and Stanley H.,  
Institutional Investors and Stock Market Volatility, 
{\em Q. J. Econ.}, 
2006, 
{\bf 121}, 461--504.

\bibitem[\protect\citeauthoryear{Glosten and Milfrom}{1985}]{Glosten:1985kx}
Glosten L. and Milgrom P., 
Bid, ask and transaction prices in a specialist market with heterogeneously informed traders, 
{\em J. Financ. Econ.}, 
1985, 
{\bf 14}, 71--100.

\bibitem[\protect\citeauthoryear{Grinold and Kahn}{2000}]{Grinold:2000bh} 
Grinold R. and Kahn R., 
{\em Active portfolio management: a quantitative approach for providing superior returns and controlling risk},  
2000 
(New York: McGraw-Hill).

\bibitem[\protect\citeauthoryear{Hasbrouck}{1991}]{Hasbrouck:1991iw}
Hasbrouck J., 
Measuring the Information Content of Stock Trade, 
{\em J. Financ.}, 
1991, 
{\bf 46}, 179--207.

\bibitem[\protect\citeauthoryear{Hasbrouck and Seppi}{2010}]{Hasbrouck:2001zr} 
Hasbrouck J. and Seppi D., 
Common factors in prices, order flows, and liquidity, 
{\em J. Financ. Econ.}, 
2001, 
{\bf 59}, 383--411.

\bibitem[\protect\citeauthoryear{Kyle}{1985}]{Kyle:1985zh}
Kyle A., 
Continuous Auctions and Insider Trading, 
{\em Econometrica}, 
1985, 
{\bf 53}, 1315--1335.

\bibitem[\protect\citeauthoryear{Lillo {\itshape{et al.}}}{2003}]{Lillo:2003vn} 
Lillo F., Farmer J. D., and Mantegna R., 
Econophysics: Master curve for price-impact function, 
{\em Nature}, 
2003, 
{\bf 421}, 176--190.

\bibitem[\protect\citeauthoryear{Lyons and Evans}{2002}]{Lyons:2002ly} 
Lyons R. and Evans M., 
Order flow and exchange rate dynamics, 
{\em J. Polit. Econ.}, 
2002, 
{\bf 110}, 170--180.

\bibitem[\protect\citeauthoryear{Marsili}{2001}]{Marsili:2001mc} 
Marsili M., 
Market mechanism and expectations in minority and majority games, 
{\em Physica A}, 
2001, 
{\bf 299}, 93--103. 

\bibitem[\protect\citeauthoryear{Marsili}{2009}]{fwmd} 
Marsili M., 
Complexity and Financial Stability in a Large Random Economy, 
2009, 
SSRN: http://ssrn.com/abstract=1415971.

\bibitem[\protect\citeauthoryear{Mastromatteo {\itshape{et al.}}}{2011}]{Mastromatteo:2011ab} Mastromatteo I., Marsili M. and Zoi P., 
Financial correlations at ultra-high frequency: theoretical models and empirical estimation, 
{\em Eur. Phys. J. B},  
2011, 
{\bf 80}, 243--253.

\bibitem[\protect\citeauthoryear{M\'{e}zard {\itshape{et al.}}}{1987}]{parisi-mezard-virasoro}
M\'{e}zard M., Parisi G. and Virasoro M. A.,  
{\em Spin Glass theory and beyond}, 
1987 
(Singapore: World Scientific).

\bibitem[\protect\citeauthoryear{Moro {\itshape{et al.}}}{2009}]{Moro:2009wy} 
Moro E,. Vicente J., Moyano L., Gerig A., Farmer J. D., Vaglica G., Lillo F. and Mantegna R., 
Market impact and trading profile of hidden orders in stock markets, 
{\em Phys. Rev. E}, 
2009, 
{\bf 80}, 066102.

\bibitem[\protect\citeauthoryear{Plerou {\itshape{et al.}}}{2002}]{Plerou:2002ve} 
Plerou V., Gopikrishnan P., Gabaix X. and Stanley H., 
Quantifying stock-price response to demand fluctuations, 
{\em Phys. Rev. E}, 
2002, 
{\bf 66}, 027104.

\bibitem[\protect\citeauthoryear{Slanina}{2008}]{slanina08} 
Slanina F., 
Critical comparison of several order-book models for stock-market fluctuations, 
{\em Eur. Phys. J. B}, 
2008, 
{\bf 61}, 225--240.

\bibitem[\protect\citeauthoryear{Torre}{1997}]{Torre:1997qf} 
Torre N.,  
{\em BARRA Market Impact Model Handbook},  
1997 
(Berkeley: BARRA Inc.).

\bibitem[\protect\citeauthoryear{T{\'o}th {\itshape{et al.}}}{2006}]{toth-kertesz}
T\'{o}th B. and Kert\'{e}sz J., 
Increasing market efficiency: Evolution of cross-correlations of stock 
returns, 
{\em Physica A}, 
2006, 
{\bf 360}, 505--515.

\bibitem[\protect\citeauthoryear{T{\'o}th, Eisler {\itshape{et al.}}}{2011}]{Toth:2011sj} 
T{\'o}th B, Eisler Z, Lillo F, Bouchaud J, Kockelkoren J. and Farmer J. D.,  
How does the market react to your order flow?, 
\texttt{arXiv:1104.0587}, 
2011.

\bibitem[\protect\citeauthoryear{T{\'o}th, Lemp{\'e}ri{\`e}re {\itshape{et al.}}}{2011}]{Toth:2011bs}
T{\'o}th B., Lemp{\'e}ri{\`e}re Y., Deremble J., de~Lataillade J., Kockelkoren J. and Bouchaud J.-P., 
Anomalous price impact and the critical nature of liquidity in financial markets, 
{\em Phys. Rev. X}, 
2011, 
{\bf 1}, 021006.

\end{thebibliography}
\end{document}